\documentclass{appolb}
\usepackage{amsmath}
\usepackage{amssymb}
\usepackage{graphicx}
\usepackage{xcolor}

\begin{document}
\title{The analytic structure of the QCD propagators, confinement, and deconfinement%
\thanks{Presented at Workshop ``Excited QCD 2026'', University of Granada, Granada (Spain), 08-14/01/2026.}%
}
\author{Giorgio Comitini
\address{Università degli Studi di Catania -- Dipartimento di Fisica e Astronomia ``E. Majorana'', and Istituto Nazionale di Fisica Nucleare (INFN) -- Sez. Catania, via Santa Sofia 64, Catania I-95123, Italy}
}
\maketitle
\begin{abstract}
We present the first complete calculation of the analytic structure of the zero-spatial-momentum finite-temperature Landau-gauge gluon propagator carried out at one loop by a massive deformation of QCD perturbation theory -- the screened massive expansion -- at temperatures ranging from $T=0$ to $T\approx 3T_{c}$. We find no signatures of deconfinement in the form of meaningful changes in said structure. We argue that, beyond Euclidean space, massive perturbative methods -- including the Curci-Ferrari model~-- might be missing crucial dynamical information as a consequence of the perturbative violation of QCD's Ward identities.
\end{abstract}
  
\section{Introduction}

Over the last fifteen years, massive perturbative approaches to infrared Quantum Chromodynamics \cite{TW11,RSTW14,SIR16a} have received widespread attention thanks to their remarkable success in reproducing the Euclidean lattice data for QCD's Landau-gauge Green's functions. Amongst their most prominent applications is the calculation of finite-temperature quantities such as the deconfinement temperature and the field correlators. The latter are especially interesting, since the two-point Green's functions of elementary fields carry information on the particle and quasi-particle spectrum of the theory, making them the go-to candidate for exploring issues of confinement and deconfinement.

Formally, the analytic nature of one-loop perturbative calculations makes it possible to easily access the complex plane of the QCD propagators, and in particular Minkowski space, where the dynamics of the theory lives. There one can look for signatures of deconfinement in the form of a change in the analytic structure of propagators. Specifically, one may expect that, in passing from the confined to the deconfined phase, the spectral functions of the temperature-dependent QCD propagators develop positive quasi-particle peaks, as is know to happen at high temperatures \cite{BP90b}, where ordinary perturbative methods~-- backed up by Hard Thermal Loop resummations -- are available, before they break down at lower temperatures.

In what follows, with the aim of looking for such signatures, we present the first complete one-loop calculation of the low-temperature analytic structure of the Landau-gauge gluon propagator at vanishing spatial momentum, as obtained by the massive perturbative approach known as the screened massive expansion.

\section{The analytic structure of the gluon propagator at finite $T$}

The screened massive expansion \cite{SIR16a,SC18} is a reorganization of the ordinary QCD perturbative series operated with the aim of treating transverse gluons as massive already at tree level, as is appropriate for gluons that dynamically acquire a mass in the deep infrared. Suppressing diagonal color indices, in a generic linear covariant gauge defined by the gauge parameter $\xi$, its zero-order Euclidean gluon propagator reads
\begin{equation}
    \Delta_{\mu\nu}^{(0)}(p)=\frac{1}{p^{2}+m^{2}}\,\left(\delta_{\mu\nu}-\frac{p_{\mu}p_{\nu}}{p^{2}}\right)+\xi\,\frac{p_{\mu}p_{\nu}}{(p^{2})^{2}}\ ,
\end{equation}
where $m^{2}$ is a gluon mass parameter. Usage of $\Delta_{\mu\nu}^{(0)}(p)$ must be compensated by a new two-point gluonic vertex $\delta\Gamma_{\mu\nu}(p)$,
\begin{equation}
    \delta\Gamma_{\mu\nu}(p)=m^{2}t_{\mu\nu}(p)\ ,
\end{equation}
usually referred to as the gluon mass counterterm, which appears in the interaction Lagrangian when $\Delta_{\mu\nu}^{(0)}(p)$ is chosen as the order zero of the expansion instead of the ordinary massless propagator.

At finite temperature $T$, the dressed Euclidean gluon propagator $\Delta_{\mu\nu}(p,T)$ is a function of the Matsubara frequency $p^{4}=\omega_{n}=2\pi n T$, $n\in\mathbb{Z}$, and of the spatial momentum ${\bf p}$. In Landau gauge ($\xi=0$) and at zero spatial momentum, $\Delta_{\mu\nu}(p,T)$ is expressed in terms of a single form factor $\Delta(\omega,T)$.

The frequency variable $\omega$ can be analytically continued to the complex plane, $\omega\to z\in\mathbb{C}$, to explore the analytic structure of the propagator at finite temperature. Thanks to the perturbative nature of the screened massive expansion, performing the analytic continuation in this framework is as easy as replacing $\omega_{n}$ with $z$ in the expressions while making sure that the branch cut generated by multiparticle thresholds in the propagator remains located in Minkowski space, $z\in i\mathbb{R}$. The temperature-dependent spectral function $\rho(\omega,T)=\frac{1}{\pi}\,\text{Im}\{\Delta(i\omega-\epsilon,T)\}$, together with the temperature-dependent complex poles $z_{k}(T)$ and their residues $R_{k}(T)$, fully determine the analytically continued propagator $\Delta(z,T)$:
\begin{equation}
    \Delta(z,T)=\sum_{k}\frac{R_{k}(T)}{z-z_{k}(T)}+\int_{-\infty}^{+\infty}d\omega\ \frac{\rho(\omega,T)}{\omega+iz}\ .
\end{equation}

In what follows, we will report on results obtained in pure Yang-Mills theory and full QCD using two sets of parameters: the first exploits the optimization of the screened massive expansion at zero temperature, the second uses fits from the lattice data. More details on the parametrization, and a thorough analysis of the results, can be found in \cite{COM26a}.

\begin{figure}[htb]
\centerline{%
\includegraphics[width=3.7cm,angle=270]{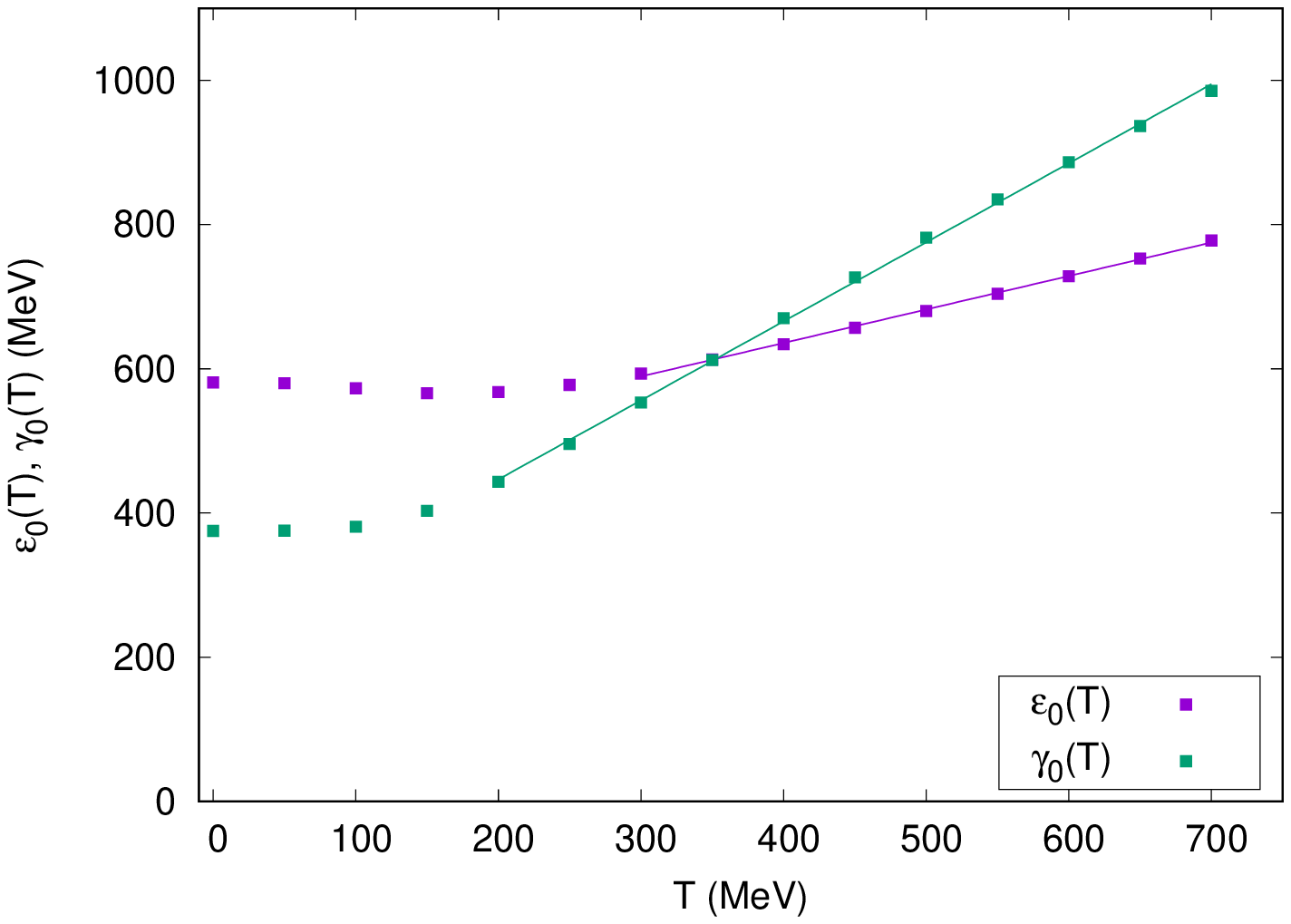}\includegraphics[width=3.7cm,angle=270]{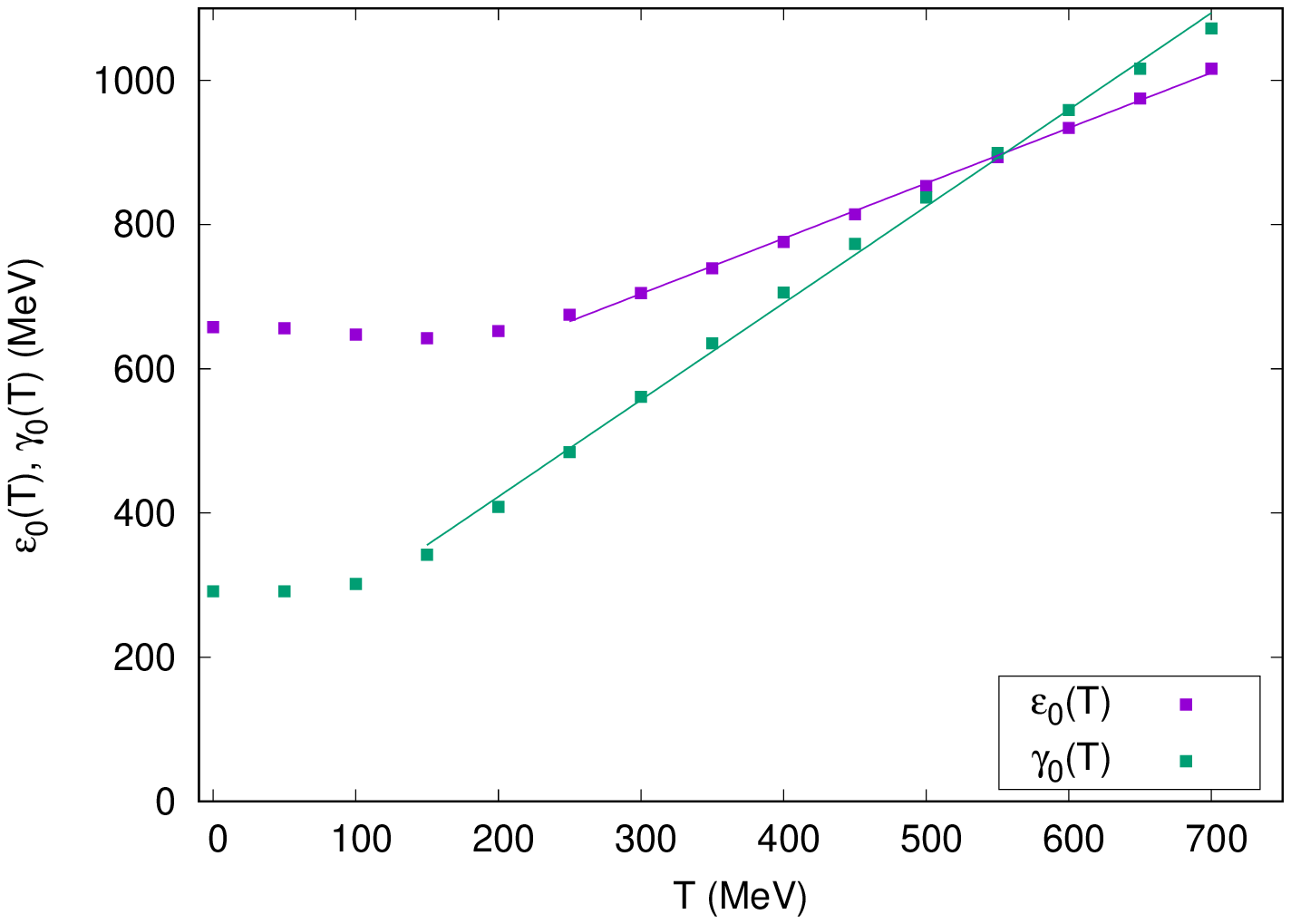}}
\centerline{%
\includegraphics[width=3.7cm,angle=270]{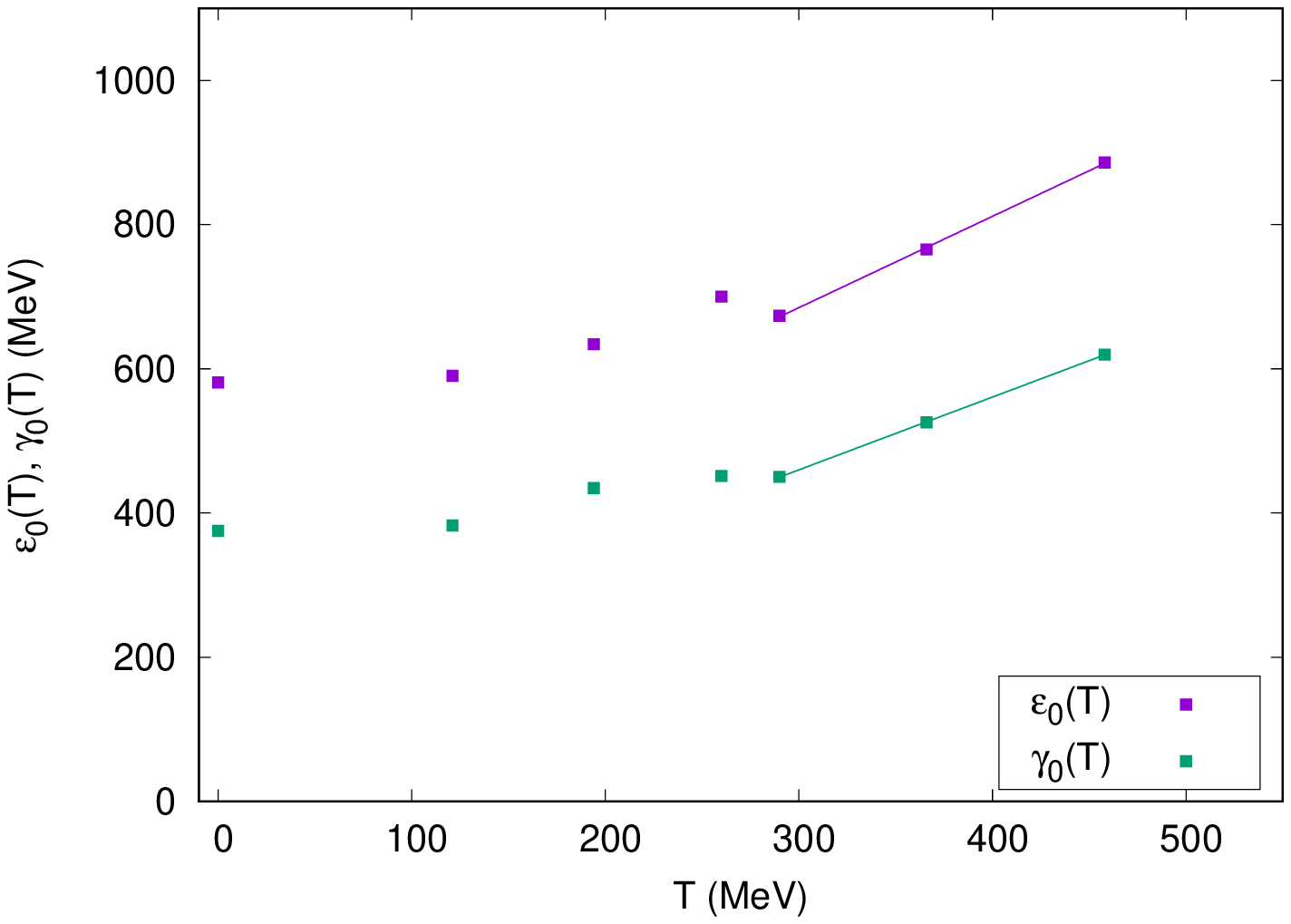}\includegraphics[width=3.7cm,angle=270]{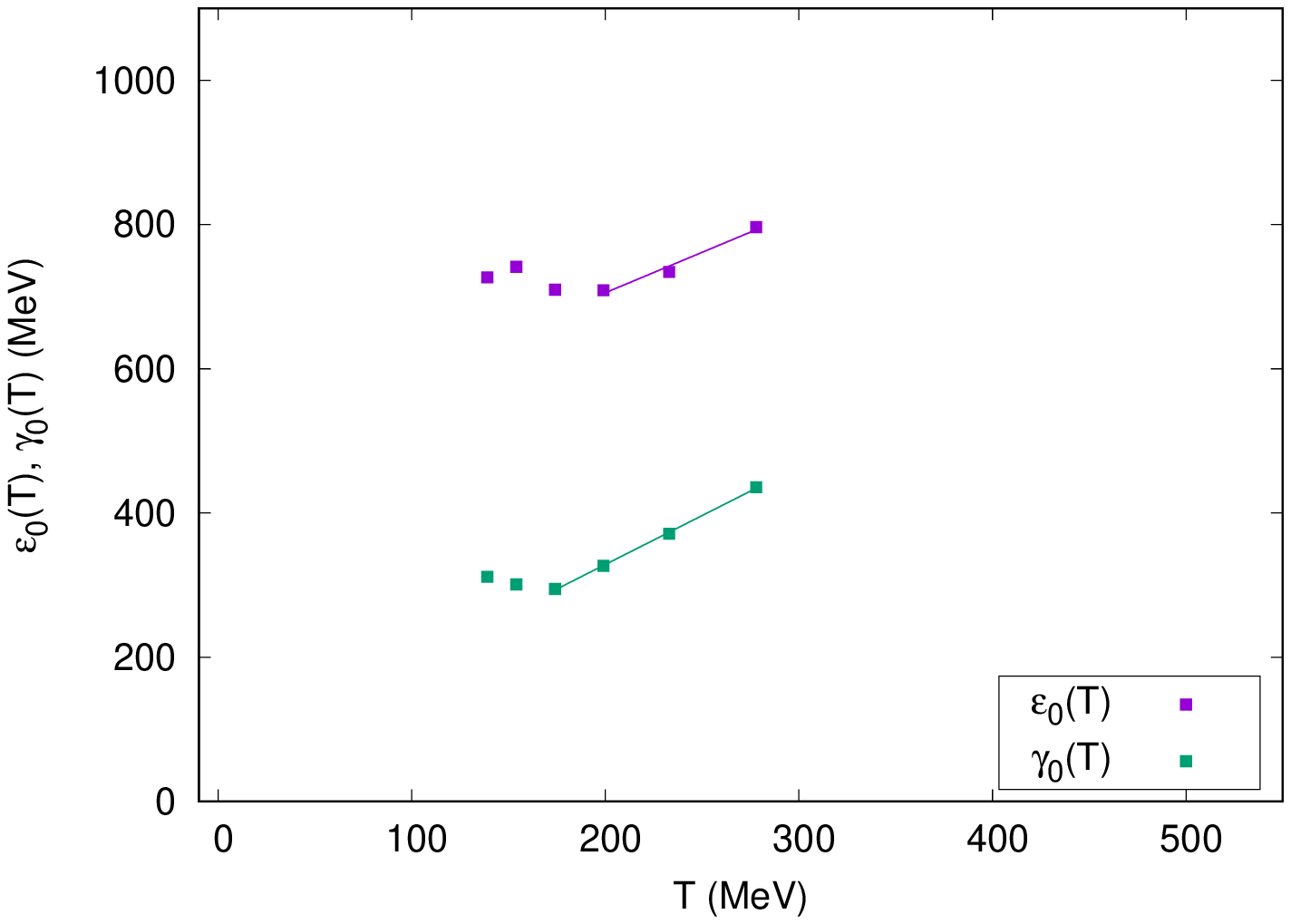}}
\caption{Real and imaginary part $\varepsilon_{0}(T)$ and $\gamma_{0}(T)$ of the poles of the finite-temperature Landau-gauge gluon propagator in pure Yang-Mills theory (left) and full QCD (right), as a function of temperature. Top: free parameters from the optimization of the screened expansion at $T=0$. Bottom: free parameters from the lattice data. $n_{F}=2+1$ on the top right, $n_{F}=2$ on the bottom right.}
\label{Fig:poles}
\end{figure}

When computed to one loop in the screened massive expansion, the Landau-gauge zero-momentum gluon propagator is found to possess a quartet of complex-conjugate poles, $z=\pm z_{0},\pm z_{0}^{\star}$ -- equivalently, a pair of complex-conjugate poles $z^{2}=z_{0}^{2},(z_{0}^{2})^{*}$ --, in its principal Riemann sheet for all explored temperatures, $T\in[0,\approx 3T_{c}]$ \cite{Sir17d}. Fig.~\ref{Fig:poles} shows the evolution with temperature of the pole with $\text{Re}(z_{0}),\text{Im}(z_{0})>0$. The real and the imaginary part of the pole become linear with $T$ at temperatures greater than $\approx T_{c}$, where the deconfinement temperature is $T_{c}\approx 270$~MeV for pure Yang-Mills theory and $T_{c}\approx 150$~MeV for full QCD. Other than this, we see no relevant change in the behavior of the poles across $T_{c}$.

\begin{figure}[htb]
\centerline{%
\includegraphics[width=3.8cm,angle=270]{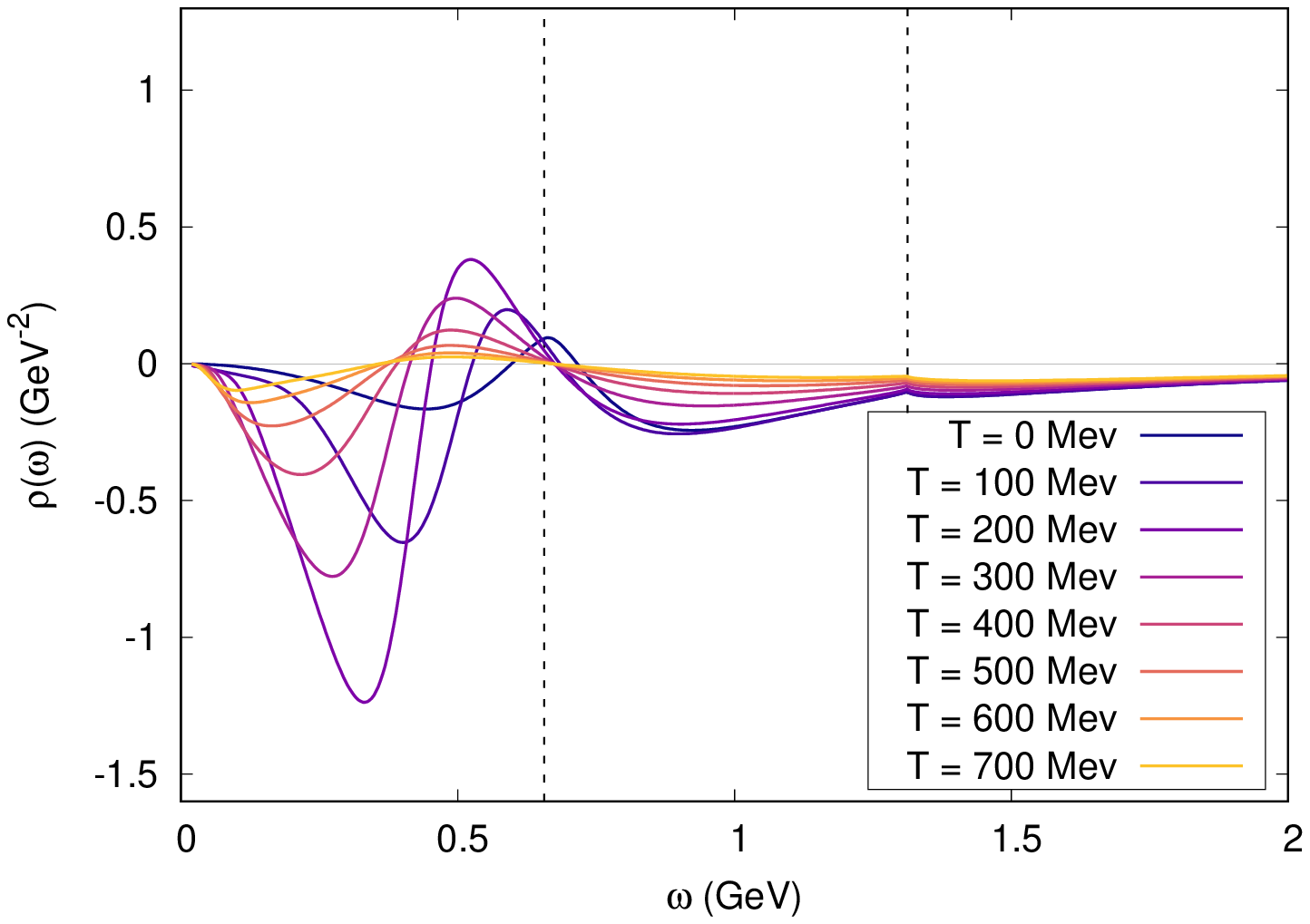}\includegraphics[width=3.8cm,angle=270]{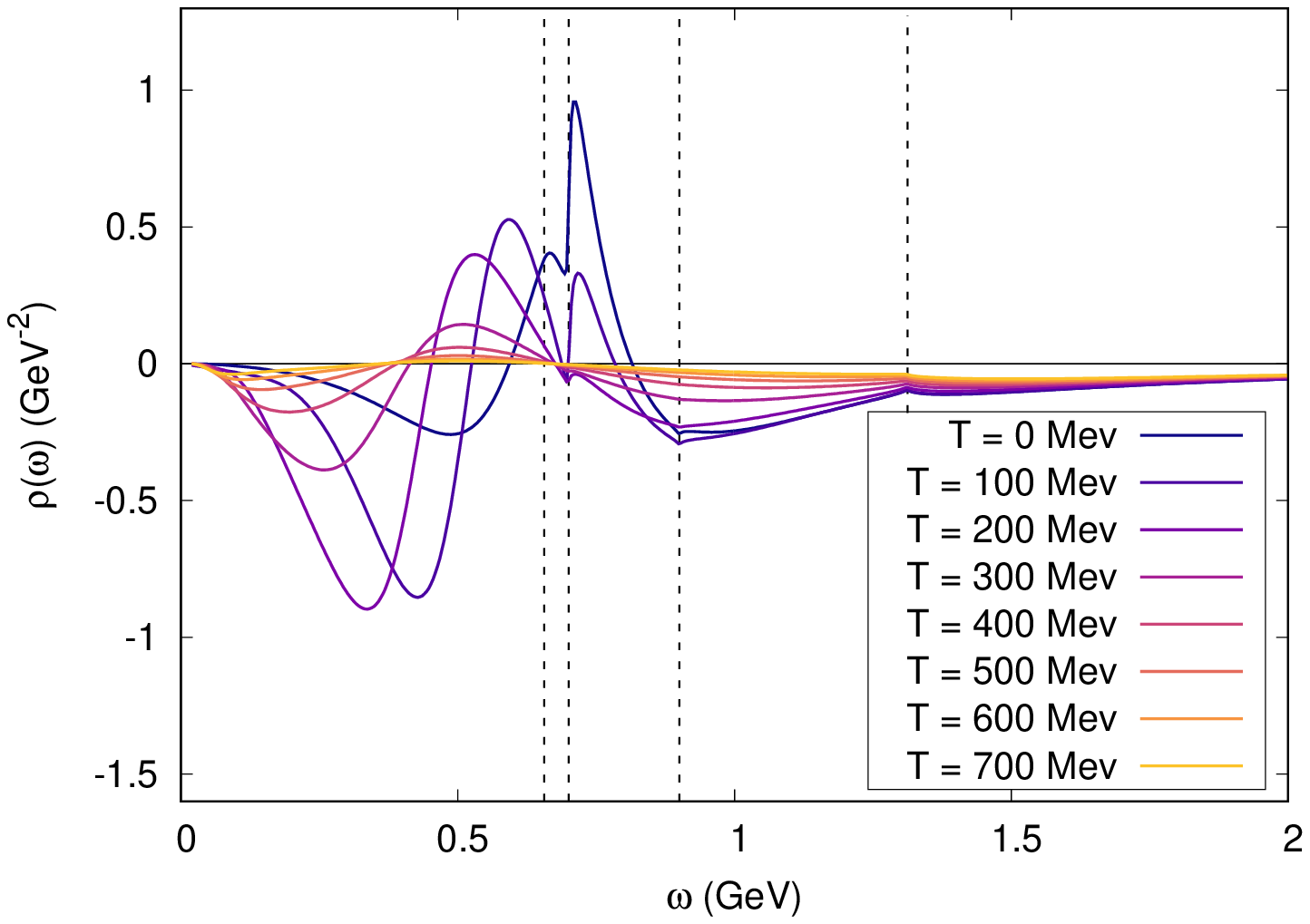}}
\centerline{%
\includegraphics[width=3.8cm,angle=270]{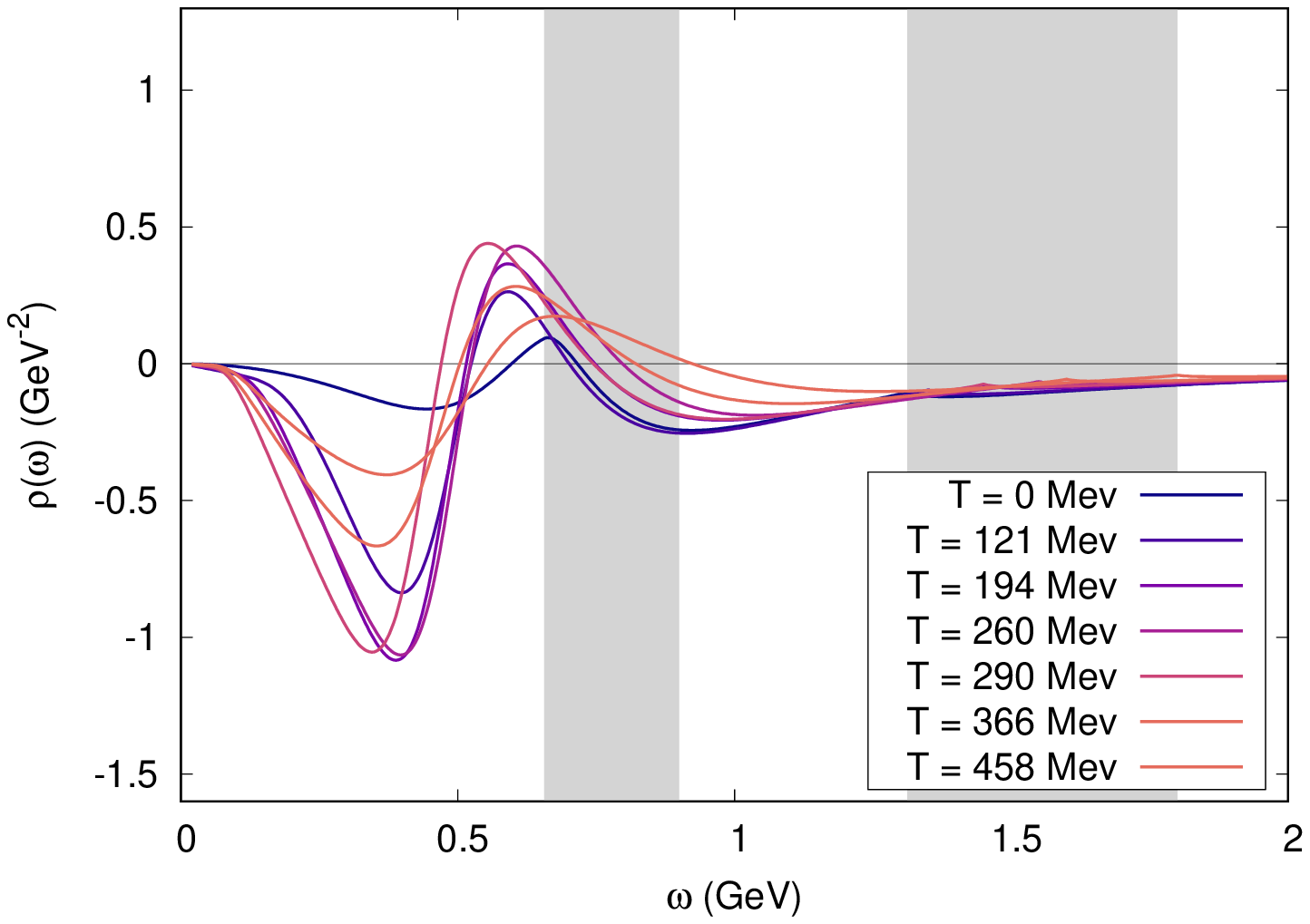}\includegraphics[width=3.8cm,angle=270]{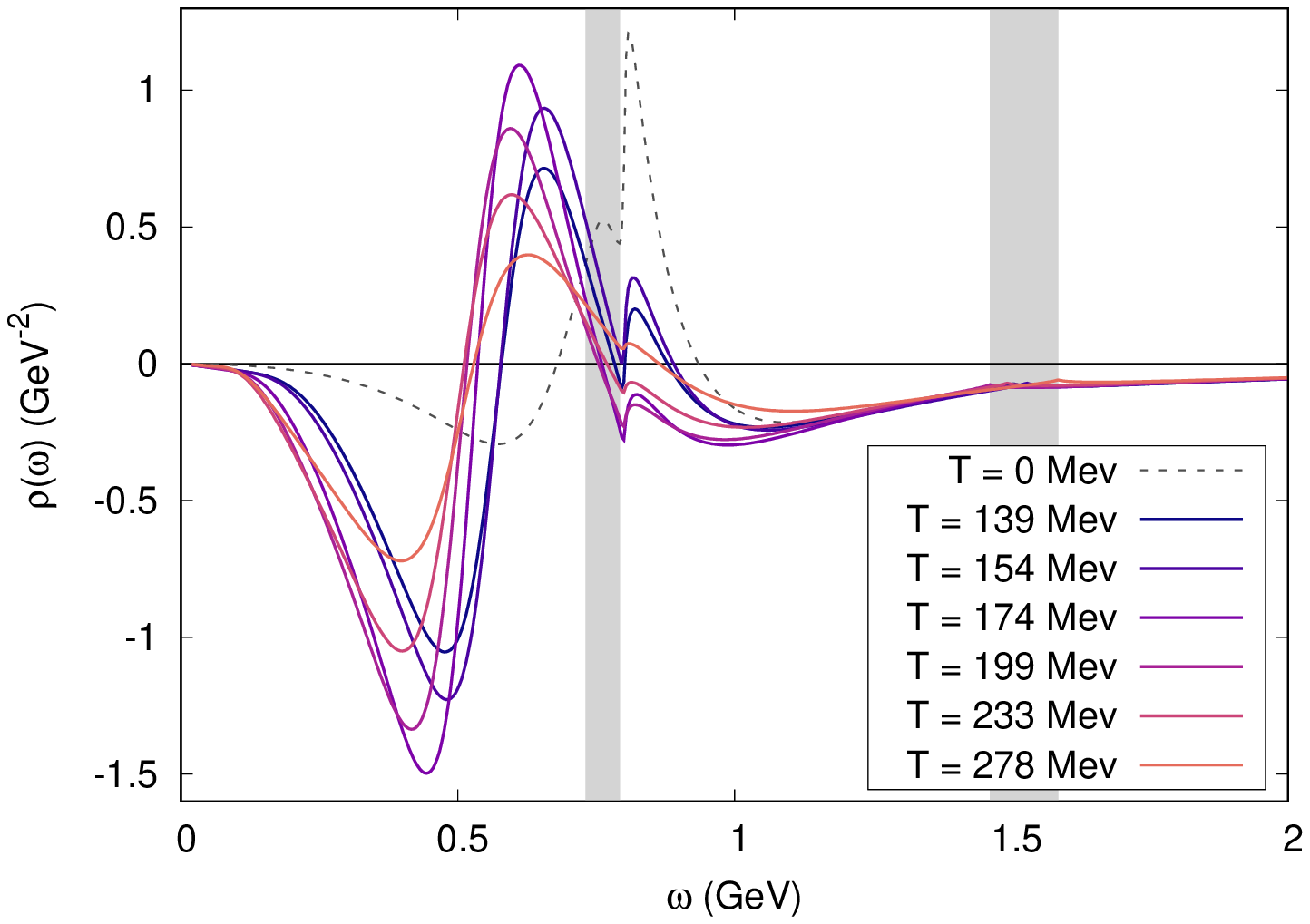}}
\caption{Spectral function of the finite-temperature Landau-gauge gluon propagator. Order of plots as in Fig.~\ref{Fig:poles}. The vertical dashed lines and grey bands represent the onset of gluon and quark mass thresholds.}
\label{Fig:spectral}
\end{figure}

In Fig.~\ref{Fig:spectral} we show the evolution of the gluon spectral function $\rho(\omega,T)$ with temperature. $\rho(\omega,T)$ displays mass thresholds at $\omega=m,2m$ in Yang-Mills theory and, additionally, at $\omega=2M$ -- $M$ being a quark mass -- in full QCD. When parameters from the lattice are used, $m=m(T)$ becomes a function of temperature, and the $\omega=m,2m$ thresholds become temperature-dependent. At all the explored temperatures, the spectral function possesses a characteristic structure of minima and maxima which was already observed in zero-temperature studies \cite{SIR16b}. In full QCD, we also observe a maximum right above the first quark mass threshold. We see no meaningful changes in the behavior of the spectral function across $T_{c}$. In particular, in the present framework, $\rho(\omega,T)$ develops no positive quasi-particle spectral peaks beyond the deconfinement transition.

\section{Complex-conjugate poles, the plasmon puzzle and QCD's Ward identities}

At high temperatures, the gluon propagator is known to develop a positive quasi-particle spectral peak, hence to behave like a massive deconfined quasi-particle, with mass $m_{g}\propto gT$ and damping $\gamma_{g}\propto g^{2}T$ \cite{BP90b}. On the other hand, it is little known that a straight one-loop calculation of the analytic structure of the gluon propagator performed in ordinary thermal perturbation theory in linear covariant gauges yields complex-conjugate poles up to arbitrarily high temperatures. Back in the 1980s, this problem was known as the plasmon puzzle \cite{Nad88}: in linear covariant gauges, the pole in the retarded gluon propagator was found to have a gauge-dependent damping $\gamma_{g}$ with the wrong sign. It is fairly easy to recognize that a retarded pole whose imaginary part has the wrong sign is nothing more than a pole in the ordinary Riemann sheet of the propagator, always accompanied by a complex-conjugate partner -- namely, the corresponding advanced pole with an equally wrongly-signed imaginary part.

The plasmon puzzle -- i.e. the appearance of gauge-dependent complex-conjugate poles in ordinary thermal perturbation theory -- was solved by the Hard Thermal Loop (HTL) resummation of Braaten and Pisarski. The structure of HTLs is rigidly fixed by gauge invariance \cite{BP90a,BP92}: the HTL in the gluon polarization is the unique transverse polarization term which is able to generate a thermal mass, whereas the HTLs in vertices are the only hard terms compatible with the Ward identities of the theory. In ordinary thermal perturbation theory, adopting a resummation scheme which -- in the presence of a thermal mass -- explicitly preserves these identities eliminates complex-conjugate poles from the gluon propagator of linear covariant gauges.

In state-of-the-art calculations like that of the present paper, massive perturbative models like the screened massive expansion and the Curci-Ferrari model employ tree-level vertices. This can be problematic for two reasons. First, already at zero temperature, using these vertices violates QCD's Ward identities: it is well known that, when gluons acquire a mass, QCD's Ward identities can only be satisfied if nonperturbative Schwinger poles appear in the vertices \cite{Cor81}. Second, at finite temperatures, disregarding effects due to lack of resummation can yield unphysical results like those described above \cite{BP90a,BP92}.

\begin{figure}[htb]
\centerline{%
\includegraphics[width=4.8cm,angle=270]{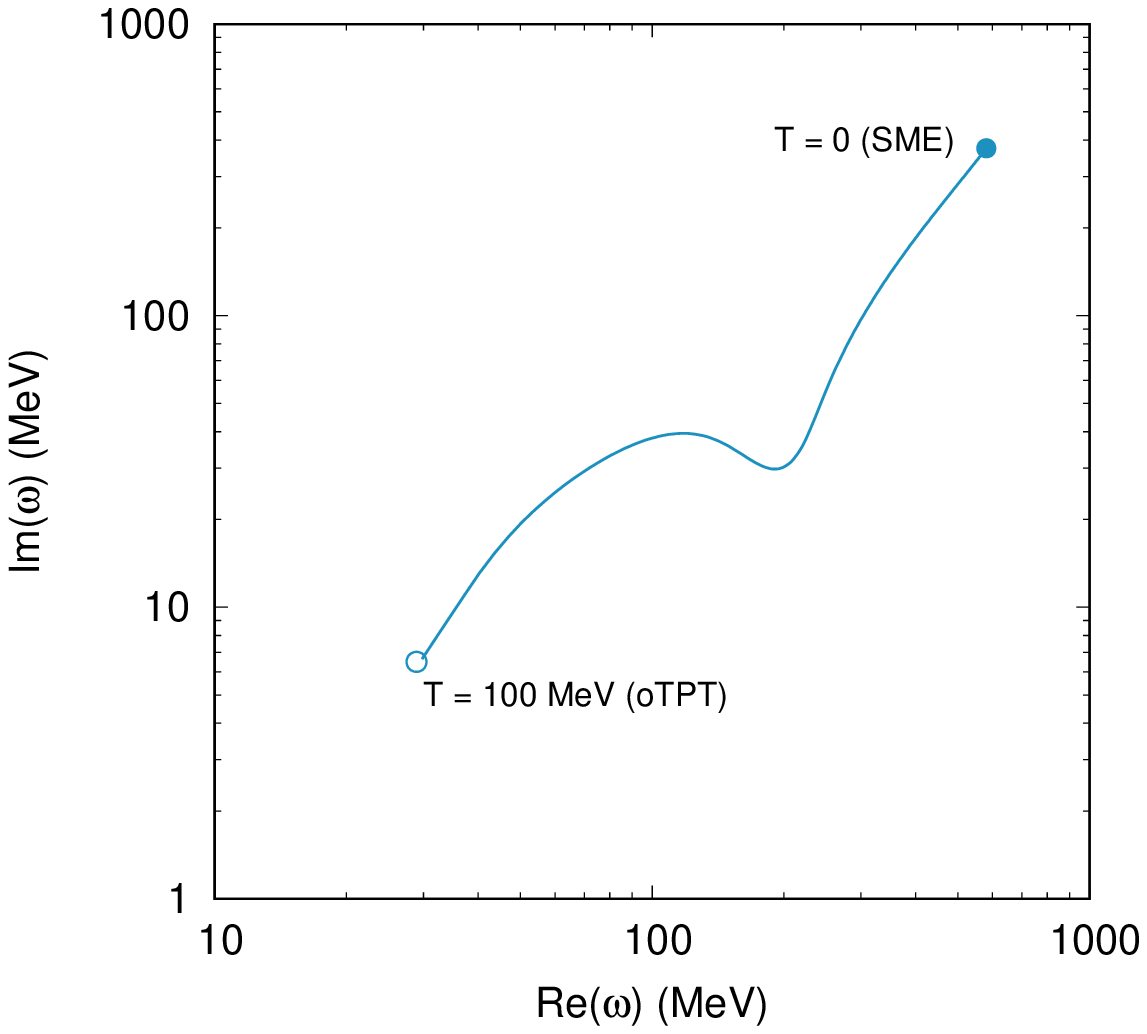}}
\caption{For $gT\gg m$ and $g\ll1$, the zero-temperature complex-conjugate poles of the screened massive expansion (SME) become the complex-conjugate poles of ordinary thermal perturbation theory's (oTPT) plasmon puzzle.}
\label{Fig:smeplasmon}
\end{figure}

The parallel between massive models and ordinary thermal perturbation theory is not only one of principle: an explicit calculation -- see Fig.~\ref{Fig:smeplasmon} -- shows that the zero-temperature complex-conjugate poles of the massive perturbative models can be continuously deformed into the complex-conjugate poles of the plasmon puzzle by tuning the temperature, coupling and gluon mass parameter. Since the finite-$T$ complex-conjugate poles of ordinary perturbation theory are known artifacts, cured by complying with QCD's Ward identities, it is of paramount importance to establish whether~-- beyond Euclidean space -- massive perturbative models are missing crucial dynamical information by disregarding the existence of nonperturbative Schwinger poles in the vertices of the theory, and by disregarding resummation effects.

\section{Conclusions}

When computed to one loop by the screened massive expansion of QCD, other than the linearization of the complex-conjugate poles' behavior with temperature, the analytic structure of the gluon propagator does not show meaningful changes beyond the deconfinement temperature. Given that, at sufficiently high temperatures, the perturbative complex-conjugate gluon poles are known artifacts caused by not complying with QCD's Ward identities in the presence of a gluon mass, it is of paramount importance to assess whether the neglect of vertex Schwinger poles and/or of resummation effects by massive perturbative models like the Curci-Ferrari model and the screened massive expansion is causing analogous artifacts down to zero temperatures.

\section*{Acknowledgments}

This work was supported by PIACERI ``Linea di intervento 1'' (M@uRHIC) of the University of Catania and by PRIN2022 (Projects No. 2022SM5YAS and No. P2022Z4P4B) within Next Generation EU fundings. Attendance to the workshop was supported by Istituto Nazionale di Fisica Nucleare (INFN), Sezione di Catania.

\bibliographystyle{unsrt}
\bibliography{QCD}

\end{document}